\documentclass[aps,prl,twocolumn,showpacs,amsmath,amssymb]{revtex4}
\usepackage{graphicx}
\usepackage{bm}

\def\br{ \mathbf{r} }
\def\bro{ \mathbf{0} }
\def\bk{ \mathbf{k} }
\def\bv{ \mathbf{v} }
\def\bqs{ \mathbf{q}_s }

\def\im{ \,{\rm Im}\,}

\begin{document}

\title{Impurity resonances in the mixed state of high-$T_c$
superconductors}

\author{K.~V.~Samokhin}

\affiliation{Department of Physics, Brock University,
St.Catharines, Ontario, Canada L2S 3A1}
\date{\today}

\begin{abstract}
We study the quasiparticle resonance states near strong impurities
in the mixed state of a $d$-wave superconductor. These states give
rise to zero-bias peaks in the local density of states, observed
in scanning tunneling microscopy experiments. The field dependence
of the peaks is obtained by averaging with respect to the
spatially non-uniform Doppler shifts in the energy of excitations.
The hybridization of the magnetic field-induced nodal
quasiparticles with the impurity resonances results in the
suppression and broadening of the zero-bias peaks. The height of
the peaks is found to scale as $(H_{c2}/H)^{1/2}$. The magnetic
field response of the peaks is shown to be strongly dependent on
the field orientation.
\end{abstract}


\maketitle

One of the striking features of high-$T_c$ superconductors
(HTSC's) is that even a single impurity has a notable effect on
the superconducting state, creating a sharp resonance state in its
vicinity, which becomes a true bound state of zero energy in the
limit of strong impurity potential (the unitary limit)
\cite{Bal95,Sal96,Sal97,Fehr96,Fla98,Zhu00}. These bound states
manifest themselves as sharp peaks near zero bias in the energy
dependence of the single-particle density of states (DoS), which
have been observed in scanning tunneling microscopy (STM)
experiments on BSCCO compound
\cite{Pan00,Hud99,Yazd99,Pan00cores}.

The bound states near non-magnetic impurities are specific to
unconventional superconductors with an anisotropic order parameter
and the gap nodes at the Fermi surface \cite{Book}. The zero width
of these states in the unitary limit follows from the absence of
hybridization with the extended quasiparticle states: HTSC
compounds are $d$-wave superconductors, and the DoS,
$N_0(\omega)$, of the extended states vanishes at small energy
linearly in $\omega$. However, this is no longer true if the
superconductor is placed in an external magnetic field $H$. The
field-induced supercurrents act as pair breakers, filling the gap
nodes and creating a finite density of bulk quasiparticles at low
energies. These quasiparticles are responsible for a number of
peculiar properties of the HTSC's, such as a non-analytical in $H$
behavior of the electronic specific heat and the thermal
conductivity in the mixed state at low temperatures (the Volovik
effect) \cite{Vol93}. One can expect that the hybridization with
the field-induced bulk quasiparticles gives rise to a finite width
of the impurity bound states. A simpler version of this problem,
applicable, for instance, to the case of $\mathbf{H}\parallel ab$
in the Meissner state, was studied in Ref. \cite{Sam02}, where a
{\em uniform} supercurrent $\bqs$ was considered and it was shown
that the bound states indeed acquire a non-zero width which
depends non-analytically on $q_s$. In this paper, we study the
effect of the magnetic field on the resonance states near unitary
impurities in the mixed state for $\mathbf{H}\parallel c$. In the
presence of vortices, the supercurrent is {\em non-uniform}, which
qualitatively changes the magnetic field response of the impurity
resonances. The unitary limit of scattering is of particular
interest because the most profound effects related to the impurity
resonances have been observed in the vicinity of Zn impurities in
BSCCO, which have the $s$-wave phase shift $\delta_0$ close to
$\pi/2$ \cite{Pan00}.

Suppose we have a repulsive point-like impurity which is described
by the potential $U(\br)=u\delta(\br)$ ($u>0$) in a
two-dimensional $d$-wave superconductor. The external magnetic
field $\mathbf{H}$ is directed along the $c$ axis. We assume that
$H_{c1}<H\ll H_{c2}$. In these conditions, the Abrikosov vortices
are well separated and the amplitude of the superconducting order
parameter is constant almost everywhere, except from the core
regions, whose size is of the order of the coherence length,
$\xi_0\sim 1$ nm. The average distance between vortices, $a_v$, is
estimated from the requirement that there is one flux quantum per
vortex: $a_v=\sqrt{\Phi_0/\pi H}$, where $\Phi_0=hc/2e$. In
typical experimental situations, $\xi_0\ll a_v\ll\lambda_L$, where
$\lambda_L$ is the London penetration depth, which allowed us to
replace the internal induction $B$ by the applied field $H$.

The quantity measured in STM experiments is the local differential
tunneling conductance, which is proportional to the local DoS,
$N(\br,\omega)=-(1/\pi)\im G^R_{11}(\br,\br;\omega)$, where $G^R$
is the retarded Gor'kov-Nambu matrix Green function. In the
presence of a single scalar impurity, one can express $G^R$ in
terms of the Green function, $G_0^R(\br_1,\br_2;\omega)$, of a
clean superconductor in the mixed state: $G^R=G^R_0+G^R_0TG^R_0$,
where $T(\omega)=u\tau_3\left[1-ug_0(\omega)\tau_3\right]^{-1}$ is
the $T$-matrix, with $g_0(\omega)=G^R_0(\bro,\bro;\omega)$. Then,
\begin{equation}
\label{Ndef}
    N(\br,\omega)=N_0(\br,\omega)+N_{imp}(\br,\omega),
\end{equation}
where $N_0$ is the local DoS for a $d$-wave superconductor in the
mixed state in the absence of impurity, and
\begin{equation}
\label{Nimp}
  N_{imp}=-\frac{1}{\pi}\im\left[G_0^R(\br,\bro;
  \omega)T(\omega)G_0^R(\bro,\br;\omega)\right]_{11}
\end{equation}
is the impurity-induced contribution. The unitary limit
corresponds to strong scattering: $u\to\infty$ and
 $T(\omega)\to-g_0^{-1}(\omega)$.
In this limit, $N(\bro,\omega)=0$ because the quasiparticles are
prevented from occupying the impurity site by a strong repulsive
potential. Thus, in order to study the effect of magnetic field on
the bound state near a unitary impurity, one should calculate the
local DoS at one of its nearest neighbors, e.g. at
$\br=\mathbf{a}$, where $N(\br,\omega)$ reaches its maximum
\cite{Sal97}. That the STM pictures show the maximum of the
tunneling conductance directly above the impurity can be
attributed to the blocking effect of Bi-O layers \cite{ZTH00} and
does not invalidate the standard theoretical model of the impurity
bound states. It should be mentioned that there are alternative
explanations of this effect. For example, it was suggested that
superconductivity is completely destroyed in the vicinity of a Zn
impurity (the ``Swiss Cheese'' model, see Ref. \cite{Nach96}),
creating an effectively normal region with high DoS. Another
possibility is the Kondo screening of the local magnetic moments
induced around the Zn impurity, see Ref. \cite{Polk00}. It is
still an open question, whether the peak in the STM conductance
right above the impurity site is related to the details of the
tunneling measurements or is an intrinsic property of the Cu-O
layers. In this article, we adopt the former point of view.

We shall see that the energy dependence of
$N_{imp}(\mathbf{a},\omega)$ is determined mostly by that of the
$T$-matrix. A closed analytical expression for
$g_0(\omega)=G^R_0(\bro,\bro;\omega)$ can be obtained only in the
case of a uniform supercurrent, when a gauge transformation can be
applied to make the order parameter real and independent of $\br$,
so that the Fourier transform of $G^R_0(\br,\bro;\omega)$ is given
by
\begin{equation}
\label{G_0uniform}
 G^R_{0,uniform}(\bk,\omega)=\frac{(\omega_+-\bv_F\bqs)\tau_0+
 \xi_\bk\tau_3+\Delta_\bk\tau_1}{(\omega_+-\bv_F\bqs)^2-
 \xi^2_\bk-\Delta^2_\bk}.
\end{equation}
Here $\omega_+=\omega+i0$, $\tau_i$ are the Pauli matrices,
$\xi_\bk$ is the normal state spectrum,
$\mathbf{v}_F=\nabla_\bk\xi_\bk$ is the Fermi velocity, $\bqs$ is
the superfluid momentum, and $\Delta_\bk=2\Delta_0(\cos k_xd-\cos
k_yd)$ is the mean-field order parameter, corresponding to
$d_{x^2-y^2}$ symmetry ($d$ is the lattice constant). We do not
calculate the order parameter self-consistently and assume
$\Delta_0$ to be constant. The numerical investigation of the
self-consistency effects shows some suppression of the order
parameter near the impurity site \cite{Hett99}, which leads only
to a renormalization of the effective impurity strength towards
the unitary limit \cite{Shnir99}. It follows from Eq.
(\ref{G_0uniform}) that the quasiparticle energy is
Doppler-shifted in the presence of supercurrent: $E_\bk=
\pm\sqrt{\xi_\bk^2+\Delta_\bk^2}+\mathbf{v}_F\bqs$.

The relevant energies of the problem are small compared to the gap
magnitude: according to Refs. \cite{Kub98}, the energy scale
associated with the Doppler shift in the mixed state is given by
$E_H=v_F/a_v\ll\Delta_0$. The Zeeman splitting is neglected
because $\mu_BH\ll E_H$ at all experimentally relevant fields
along the $c$ axis (the field at which the Zeeman splitting
becomes comparable to $E_H$ is of the order of $H_p^2/H_{c2}$,
where $H_p$ is the paramagnetic limiting field). Assuming an
electron-hole symmetric band, the momentum integrals can be easily
calculated, giving the following result at real frequencies:
$g_0(\omega)=\pi N_FF(z)\tau_0$, where $z=\omega/\Delta_0$ and
\begin{equation}
\label{F_def}
   F(z)=\sum\limits_{n=1}^4\left[\frac{1}{2\pi}(z-z_n)
   \ln|z-z_n|-\frac{i}{4}|z-z_n|\right].
\end{equation}
Here $N_F$ is the DoS in the normal state at the Fermi level, $n$
labels the gap nodes, $z_n={\bf v}_n\bqs/\Delta_0$ are the
dimensionless Doppler shifts, and $\bv_n$ are the Fermi velocities
at the nodes (we label the nodes in such a way that $z_1=-z_3$,
and $z_2=-z_4$). Expression (\ref{F_def}) is valid at $|z|\ll 1$,
$|z_n|\ll 1$. Finally, $T(z)=uc\tau_3[c-F(z)\tau_3]^{-1}$, where
$c=1/(\pi uN_F)=\cot\delta_0>0$ for a repulsive impurity.

The spectrum of the bound states is determined by the poles of the
$T$-matrix, i.e. by the equation $F(z)=\pm c$. In zero field,
$F(z)\to F_0(z)=(2/\pi)z\ln|z|-i|z|$. At $c\ll 1$, the equation
$F_0(z)=\pm c$ has the solution $z_0=\mp\pi c/(2|\ln
c|)-i\pi^2c/(4\ln^2c)$ with logarithmic accuracy, describing a
narrow impurity-induced resonance \cite{Bal95}. In the unitary
limit $c=0$, the resonance is decoupled from the continuum of bulk
excitations and becomes a true bound state of zero width, which
manifests itself as a sharp peak in the tunneling DoS. In the
presence of a non-zero supercurrent, the DoS of the bulk
excitations does not vanish at $\omega=0$, which leads to a
stronger hybridization of the bound state and the continuum of
propagating states. As a result, the bound state is replaced by a
sharp resonance whose width is proportional to
$v_Fq_s/|\ln(v_Fq_s/\Delta_0)|\ll\Delta_0$ \cite{Sam02}.

In the mixed state, the supercurrent and the order parameter
amplitude are non-uniform, and Eq. (\ref{G_0uniform}) is no longer
valid. Further progress can be achieved by using the so-called
Doppler-shift approximation \cite{Vol93,Kub98}. The basic
assumption of this approach that the low-energy properties of a
superconductor with the gap nodes are determined by the extended
quasiparticle states and the contribution from the vortex cores
can be neglected. The effect of the magnetic field can be
described quasiclassically by introducing a non-uniform Doppler
shift of the excitation energy: $E_\bk\to
E_\bk(\br)=E_\bk+\mathbf{v}_F\bqs(\br)$. In particular, the Green
function at coinciding arguments becomes
$G_0^R(\br,\br;\omega)=\pi N_FF(z)\tau_0$, where $F$ is defined by
Eq. (\ref{F_def}), in which the Doppler shifts $z_n$ now depend on
the coordinate: $z_n\to z_n(\br)={\bf v}_n\bqs(\br)/\Delta_0$. It
is expected that the Doppler-shift approach works well at low
fields $H\ll H_{c2}$ (for a discussion of its applicability and
limitations, see, e.g., Ref. \cite{Vekh01}).

The Doppler-shift approximation should be used with caution in the
calculation of the local DoS because the impurity-induced
contribution rapidly oscillates as a function of $\br$ with the
period of the order of $k_F^{-1}$. We assume that the impurity is
unitary, so that
\begin{eqnarray}
\label{Nunitary}
    N(\br,\omega)=N_0(\br,\omega)\nonumber+
    \frac{1}{\pi^2N_F}\im\Bigl\{F^{-1}\left(\frac{\omega}{\Delta_0}\right)\\
    \times\left[G_0^R(\br,\bro;\omega)G_0^R(\bro,\br;\omega)\right]_{11}
    \Bigr\}.
\end{eqnarray}
Due to the smallness of the energy scale $E_H$, only the limit
$(\omega,v_Fq_s)\leq E_H\ll\Delta_0$ is relevant. In this limit,
one can neglect $N_0(\br,\omega)$ compared to
$N_{imp}(\br,\omega)$, which is peaked at small $\omega$ because
of the singularity of $F^{-1}(\omega/\Delta_0)$. In addition, the
product of two Green functions in the second term on the
right-hand side of Eq. (\ref{Nunitary}) depends weakly on $\omega$
and $q_s$ and can therefore be replaced by its value at
$\omega=q_s=0$:
\begin{equation}
    \left[G_0^R(\br,\bro;0)G_0^R(\bro,\br;0)\right]_{11}=
    (\pi N_F)^2\gamma(\br).
\end{equation}
Here $\gamma=|I_1|^2+|I_2|^2$ is a real number, which depends on
the order parameter and the shape of the Fermi surface,
\begin{eqnarray*}
    I_1(\br)=\frac{1}{\pi
    N_F}\int\frac{d^2\bk}{(2\pi)^2}\;e^{i\bk\br}
    \frac{\xi_\bk}{\xi_\bk^2+\Delta_\bk^2},\\
    I_2(\br)=\frac{1}{\pi
    N_F}\int\frac{d^2\bk}{(2\pi)^2}\;e^{i\bk\br}
    \frac{\Delta_\bk}{\xi_\bk^2+\Delta_\bk^2},
\end{eqnarray*}
where the $\bk$-integration goes over the first Brillouin zone
$-\pi/d\leq k_x,k_y\leq \pi/d$. Thus, in the unitary limit and at
$\omega\ll\Delta_0$,
\begin{equation}
    \frac{N_{imp}(\mathbf{a},\omega)}{N_F}=\gamma(\mathbf{a})\im F^{-1}
    \left(\frac{\omega}{\Delta_0}\right).
\end{equation}
For a dispersion $\xi_\bk=-2t(\cos k_xd+\cos k_yd)-4t_1\cos
k_xd\cos k_yd -\mu$, and $\br=\mathbf{a}=\hat xd$, we have
$\gamma(\mathbf{a})=0.63$ (we used the parameters representative
of BSCCO near the optimum doping: $t_1/t=-0.3$, $\mu/t=-1$,
$\Delta_0/t=0.04$, $\Delta_0=10$ meV).

When calculating the energy dependence of the impurity resonance
contribution to the local DoS,
$N_{imp}(\omega)=N_{imp}(\mathbf{a},\omega)$, one can use the
values of the Doppler shifts at the impurity site,
$z_n=z_n(\bro)$, since the variation of the supercurrent at the
atomic length scale is negligibly small. As a final step, one
should average over all impurity positions (or, equivalently, over
the vortex positions relative to the impurity site):
\begin{equation}
\label{averageDoS}
    \frac{N_{imp}(\omega)}{N_F}=\gamma\left\langle\im
    F^{-1}\left(\frac{\omega}{\Delta_0}\right)\right\rangle_{\cal P}.
\end{equation}
Here $\langle\im F^{-1}\rangle_{\cal P}=\int dz_1\,dz_2\,{\cal
P}(z_1,z_2)\im F^{-1}$, where $F$ is defined in Eq. (\ref{F_def}),
and $\cal P$ is the distribution function of the Doppler shifts:
\begin{eqnarray}
\label{P}
    {\cal P}(z_1,z_2)=\frac{1}{{\cal A}}\int d^2\br\;
    \delta\left(z_1-\frac{\bv_1\bqs(\br)}{\Delta_0}\right)\nonumber\\
    \times\delta\left(z_2-\frac{\bv_2\bqs(\br)}{\Delta_0}\right),
\end{eqnarray}
where ${\cal A}$ is the system area.

According to Ref. \cite{Pan00cores}, Zn impurities in BSCCO act as
pinning centers for the vortices, so that a sizeable fraction
($\sim 50\%$) of the vortices reside at the impurity sites,
creating a disordered vortex solid. Because of a strong variation
of the order parameter in the vortex core, the Doppler-shift
approach is clearly inapplicable for the bound states near such
impurities.  On the other hand, in the experimental conditions of
Ref. \cite{Pan00cores}, the majority of Zn atoms are found in the
space between the vortex cores and exhibit zero-bias peaks in the
tunneling DoS. The field dependence of these peaks can be analyzed
using the Doppler-shift approximation. The overlap of the impurity
resonances and the core states is neglected.

\begin{figure}
\includegraphics[width=7.6cm]{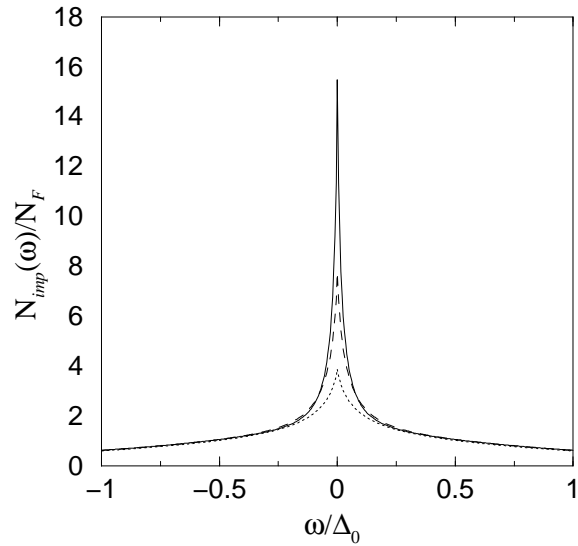}
\caption{\label{Fig1} The suppression of the impurity contribution
to the local DoS at the nearest-neighbor site $\br=\mathbf{a}$, at
different values of the magnetic field: $z_H=0.1$ (solid line),
$z_H=0.2$ (dashed line), and $z_H=0.4$ (dotted line). }
\end{figure}

The probability distribution, Eq. (\ref{P}), of the Doppler shifts
at the inequivalent gap nodes depends on the vortex configuration.
In the case of a disordered vortex solid, which is relevant to the
experiment, the probability density is peaked around the zero
Doppler shift. In Ref. \cite{Vekh01}, different models for the
supercurrent distribution were discussed giving rise to different
${\cal P}(z_1,z_2)$. One possible choice, which agrees with the
results of numerical simulations and is believed to capture the
physics reasonably well, is given by
\begin{equation}
\label{Plattice}
    {\cal P}(z_1,z_2)=
    \frac{1}{\pi}\frac{z_H^2}{(z_1^2+z_2^2+z_H^2)^2},
\end{equation}
where $z_H=E_H/\Delta_0\simeq\sqrt{H/H_{c2}}$. We use this
distribution in Eq. (\ref{averageDoS}) to calculate the average
DoS.

At zero field, $z_H=0$, and ${\cal
P}(z_1,z_2)=\delta(z_1)\delta(z_2)$. In this case, the local DoS
has a divergent peak near zero bias:
$N_{imp}(\omega)/N_F=\gamma\im F_0^{-1}(z)$ [it is not a
$\delta$-function peak because $F_0^{-1}(z)$ does not have a
simple pole structure]. At finite $H$, the height of the peak at
$\omega=0$ becomes finite and can be calculated exactly:
\begin{eqnarray}
\label{zero_bias}
    \frac{N_{imp}(0)}{N_F}&=&2\gamma\left\langle
    \frac{1}{|z_1|+|z_2|}\right\rangle_{\cal P}\nonumber\\
    &=&\frac{2\sqrt{2}\ln(\sqrt{2}+1)\gamma}{z_H}\propto
    \sqrt{\frac{H_{c2}}{H}}.
\end{eqnarray}
For the band dispersion described above, $N_{imp}(0)/N_F\simeq
1.57\sqrt{H_{c2}/H}$. This should be contrasted to the uniform
supercurrent model considered in Ref. \cite{Sam02}, according to
which $N_{imp}(0)/N_F\propto H^{-1}$.

It can also be shown that the result $N_{imp}(0)/N_F\propto
\sqrt{H_{c2}/H}$ is actually independent of the choice of ${\cal
P}(z_1,z_2)$. Indeed, ${\cal P}(z_1,z_2)$ is even in both $z_1$
and $z_2$ and also symmetric with respect to the interchange
$z_1\leftrightarrow z_2$. If the only energy scale of the Doppler
shift distribution is $E_H$, then ${\cal P}$ can always be written
in the form
\begin{equation}
    {\cal P}(z_1,z_2)=\frac{1}{z_H^2}\;
    f\left(\frac{z_1}{z_H},\frac{z_2}{z_H}\right),
\end{equation}
where $f(x,y)$ is a dimensionless function such that
$f(x,y)=f(y,x)$, $f(\pm x,\pm y)=f(x,y)$, and
$\int_{-\infty}^\infty dx\,dy\,f(x,y)=1$ in order to satisfy the
normalization condition for ${\cal P}$. Then,
\begin{eqnarray}
    \frac{N_{imp}(0)}{N_F}=\frac{2\gamma}{z_H^2}\int
    \frac{dz_1\,dz_2}{|z_1|+|z_2|}\;
    f\left(\frac{z_1}{z_H},\frac{z_2}{z_H}\right)\nonumber\\
    =A\frac{\Delta_0}{E_H},
\end{eqnarray}
where $A=2\gamma\int dx\,dy\,f(x,y)/(|x|+|y|)$.

\begin{figure}
\includegraphics[width=7.6cm]{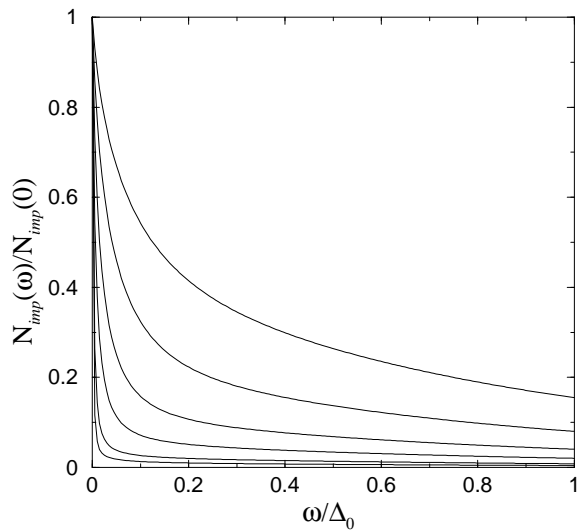}
\caption{\label{Fig2} The broadening of the DoS peak at
$\br=\mathbf{a}$ caused by magnetic field. The normalized average
DoS is calculated at different magnetic fields: $z_H=0.01$ (bottom
curve), $z_H=0.02$, $z_H=0.05$, $z_H=0.1$, $z_H=0.2$, $z_H=0.4$
(top curve). Since the peaks are symmetric with respect to
$\omega\to-\omega$, only the positive bias is shown.}
\end{figure}

At arbitrary bias, the impurity contribution to the local DoS
given by Eq. (\ref{averageDoS}) can be calculated numerically. In
Fig. \ref{Fig1}, we plot $N_{imp}(\omega)/N_F$ for several values
of $z_H$. The most prominent feature of these graphs is that the
peaks are indeed quickly suppressed as $H$ grows, in accordance
with the result (\ref{zero_bias}). To see that this suppression is
accompanied by broadening and to facilitate comparison of the
results at different $z_H$, we plot in Fig. \ref{Fig2} the local
DoS normalized to its maximum value, $N_{imp}(\omega)/N_{imp}(0)$.
The broadening of the peaks can be attributed to the hybridization
of the impurity bound states with the field-induced nodal
quasiparticles, whose density of states at zero energy increases
with field as $\sqrt{H/H_{c2}}$ \cite{Vol93}. One should keep in
mind that since the expressions (\ref{F_def}) and
(\ref{averageDoS}) are applicable only at $z,z_H\ll 1$, they give
a quantitatively correct picture only in the vicinity of the
zero-bias peaks at fields far from $H_{c2}$.

In conclusion, we have studied the influence of magnetic field on
the impurity bound states in $d$-wave superconductors. The
magnetic field effects are treated semi-classically, using the
Doppler-shift approach. The main result is that the bound states
are destroyed by magnetic field, which manifests itself as the
suppression and broadening of the zero-bias peaks in the tunneling
DoS. The height of the peaks scales as $H^{-1/2}$ for any
non-uniform supercurrent distribution, depending on a single
energy scale $E_H$. This result is sensitive to the orientation of
the field: if $H$ is parallel to the $ab$ planes, then the uniform
supercurrent model of Ref. \cite{Sam02} should be more
appropriate, which predicts that the height of the peaks is
proportional to $H^{-1}$. As mentioned above, the impurity
resonance model is not the only possible explanation of the
zero-bias anomalies in the tunneling conductance of HTSC's. It
would be interesting to look at the magnetic-field response in the
other models, which might help to resolve the controversy about
the origin of
the zero-bias peaks.\\

The author thanks B. Mitrovi\'c for valuable advices and E.
Zijlstra for the help with numerical calculations. This work was
supported by the Natural Sciences and Engineering Research Council
of Canada.

\end{document}